\documentclass[submission,copyright,creativecommons]{eptcs}
\providecommand{\event}{TAV-WEB 2010} 
\usepackage{breakurl}             

\usepackage{epsfig, color, subfigure}
\usepackage{setspace}

\title{Contracting the Facebook API\thanks{This work is supported by NSF grants CCF-0916112 and
CCF-0716095.}
}
\author{Ben Rubinger and Tevfik Bultan
\institute{Computer Science Department\\
University of California\\
Santa Barbara, CA 93106, USA}
\email{\{brubinger,bultan\}@umail.ucsb.edu}
}
\def\titlerunning{Contracting the Facebook API}
\def\authorrunning{B. Rubinger \& T. Bultan}
\begin{document}

\maketitle

\begin{abstract}
In recent years, there has been an explosive growth in the popularity
of online social networks such as Facebook. In a new twist, third party
developers are now able to create their own web applications which plug into
Facebook and work with Facebook's ``social'' data, enabling the entire Facebook user
base of more than 400 million active users to use such applications. 
These client applications can contain subtle errors that can be hard to
debug if they misuse the Facebook API. In this paper we present an
experience report on applying Microsoft's new code contract system 
for the .NET framework to the Facebook API. We wrote contracts for several
classes in the Facebook API wrapper which allows Microsoft .NET developers
to implement Facebook applications. We evaluated the usefulness of these
contracts during implementation of a new Facebook application. Our experience
indicates that having code contracts provides a better and quicker software
development experience.
\end{abstract}

\section{Introduction}

Recently, there has been a significant advancement in the proliferation
of the social networking phenomenon. Websites such as Facebook.com
allow users to connect in new ways: sharing information with one another
and creating personal online ``social networks'' by designating other
users of the site as online friends.  

As of May, 2007, software developers can develop their own applications for
Facebook. These applications allow software developers to provide users of
their applications functionality beyond what the original Facebook website
offers. Just as the original Facebook website does, these third party
applications can take advantage of the social graph (the collection of
friendship relationships specific to each user) to provide exciting and
novel user experiences.

Examples of third party Facebook applications that currently exist include
applications that allow users to share music tastes with their friends,
give custom questionnaires/polls to their friends, allow their friends
to leave them anonymous messages, share information about the locations
where they have traveled, and more. The common theme of these applications
is that they allow interaction and observation across the social axis
by accessing the friendship relationships of users that are stored
by Facebook.

There is a huge potential user base for third party Facebook applications.
Facebook reports that they have 400 million active users, of which 50\%
sign in daily. Hence, there are many software developers writing third party
Facebook applications for this market.  There are often several different
applications performing similar functions competing for market share.
For these applications to be successful, the user experience needs to
be positive.  Facebook has built in mechanisms which let users know what
applications their friends use. Furthermore, there is a Facebook application
directory  where end users have the ability to give specific applications
ratings and to see statistics about application usage.  

A third party Facebook application that is not dependable is unlikely
to be successful.  Yet it is challenging to write bug-free third party
Facebook applications. Third party Facebook applications use the Facebook
API to access the friendships relationships among users.  An incorrect
call to a Facebook method can cause a fatal error for any third party
application. Hence, a crucial factor in writing dependable third party
Facebook applications is the correct usage of the Facebook API.  However,
the behavior of Facebook API methods are mainly described in English in
the comments provided in the Facebook API. Such informal specifications can
be misunderstood and can lead to errors. Moreover, these errors can be hard to
find and debug since there is no automated way of tracking such violations.

In this paper we investigate application of the code contracts to
the Facebook API to improve the dependability of third party Facebook
applications.
Code contracts allow developers to specify method preconditions,
postconditions and class invariants in a way that is verifiable at runtime
and in some cases statically at compile time.

Design by contract has been an influential concept in software engineering in
the last two decades~\cite{Mey92}. Many tools have been developed for analyzing
contracts. Some of these tools use the host language expressions to write
contract clauses and support runtime monitoring of contracts~\cite{KA05}.
Some tools require the use of a separate annotation language such as
Java Modeling Language (JML) or Spec\# for writing contracts~\cite{BCC05,BDF05}.
In addition to runtime monitoring, there has also been significant amount of
work on static verification of contracts~\cite{FLL02}.
There has also been work on automatically inferring contracts by observing
executions of a system~\cite{NE02}, although a combination of automatically inferred contracts and
hand written contracts seems to be the most effective approach~\cite{PCM09}.

Code contracts can be especially useful for third party application developers.
They can serve as a contract between the organization that provides the API and the
third party developers that use it. 
Code contracts can be used to specify the conditions that a third party
application has to establish on the arguments that are passed to an API method (i.e., pre-conditions)
and the conditions that the API method is expected to establish (i.e., post-conditions).
Expressing these conditions using code contracts has many advantages:
\begin{itemize}
\item The API contract would provide an unambiguous specification of how to use the API
instead of the potentially ambiguous descriptions provided in comments.
\item Using contract monitoring tools the API contract conditions can be tracked at runtime. This
would provide a useful testing tool for finding bugs in third party applications.
\item Using static analysis tools, some classes of contract violations can be detected
statically, eliminating bugs early before runtime.
\item The API contract would help in debugging third party applications
by identifying causes of the bugs and where to
assign the blame in case of a contract violation.
\end{itemize}

In this paper we present an experience report on applying code contracts to the
Facebook API. In particular we used the code contract system provided 
by the Visual Studio 2010 (the integrated development environment for doing .NET development) 
and wrote contracts for the Facebook API using this code contract system.
We have focused on several classes in the Facebook API that provides access to social
data stored by Facebook. We also conducted a case study on a Facebook application
to evaluate the effectiveness of code contracts in third party Facebook application development.
The Facebook application we developed is called Pacebook and it enables users to log their runs,
review their prior runs and 
share them with their friends. 
Our experience in using code contracts in development of Pacebook and for the Facebook API
suggests that code contracts are useful for third party application developers since they
help both in identifying bugs and removing them quickly.

The rest of the paper is organized as follows: 
In Section~\ref{pacebook} we described the Facebook application that we used
as a case study.
In Section~\ref{api} we give an overview of the Facebook API
and in Section~\ref{contract} we give an overview of the Microsoft code contract system that
we used.
In Section~\ref{api-contract} we describe how we applied code contracts to the 
Facebook API and in Section~\ref{discussion} we discuss our results.
Finally, in Section~\ref{conclusion} we conclude the paper.

\section{A Facebook Application for Run Logging}
\label{pacebook}

Pacebook is a third party Facebook application 
that allows Facebook users to keep
track of their runs/jogs by logging them and by reviewing past
entries. In addition, users are able to view information about their Facebook
``friends.'' Users can log details such as the date of the run, the
distance of the run and the duration of the run.  Friends (both Pacebook
users and other Facebook users) are able to then post comments about runs that their
friends post. Users are also able to record other friends of theirs who joined
them on their runs. Facebook refers to this process of marking others who
are in content as ``tagging.'' (Often this is done to mark individuals who
appear in photos.) If another user who was tagged in a run is not currently
a user of Pacebook, the logging still takes place on the user's behalf.

\begin{figure}[t]
\centering
\includegraphics[width=1\linewidth]{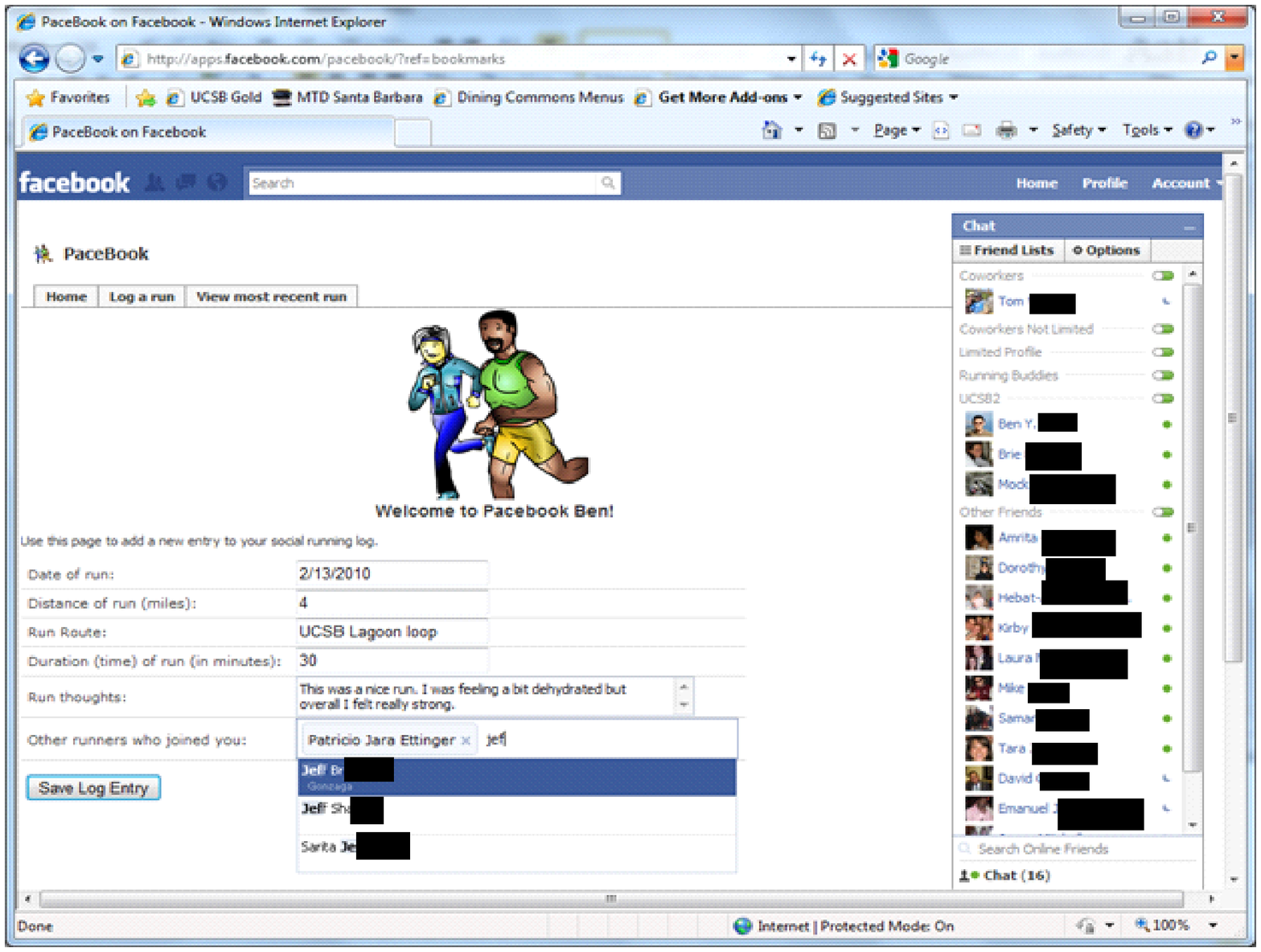}
\caption{Pacebook: A Facebook Application}
\label{screen1}
\end{figure}

When the user logs into Facebook, a link to the Pacebook application
appears on the right-hand side of the screen. Clicking this application
link launches PaceBook in the context of Facebook, providing an integrated
social experience. The screenshot in Figure~\ref{screen1} shows the interface for
logging a run. (Note that in the following screenshots, last names have
been redacted for privacy reasons.)

There are several noteworthy points about the screenshot shown in Figure~\ref{screen1}:
\begin{itemize}
\item {\em The address of the application:} The address being pointed to in the address bar points to the
facebook.com domain (http://apps.facebook.com) even though this is a
third party application. This is interesting because the third party
application is hosted on a separate web server, not on Facebook's
servers. After an initial request is made to the Facebook site, another
request is made to the third party web server to pull the application
content.
\item {\em The chat content on the right hand side:} Facebook provides a chat
facility using CSS styling to simulate a chat window which is prevalent
throughout the entire Facebook session, including while using Pacebook. Users
can engage in real-time online chat during their Pacebook experience if
they choose.
\item {\em The friend selector toward the bottom of the screenshot:} This is
achieved with the Facebook multi-friend input. This allows the user to tag
his or her friends in the run. The names with which it is populated come from
the user's Facebook friend list, and it features an autocomplete feature
which uses AJAX to dynamically populate suggestions as the user is typing. This
provides a user experience that is consistent with other Facebook features.
\end{itemize}

\begin{figure}[t]
\centering
\includegraphics[width=1\linewidth]{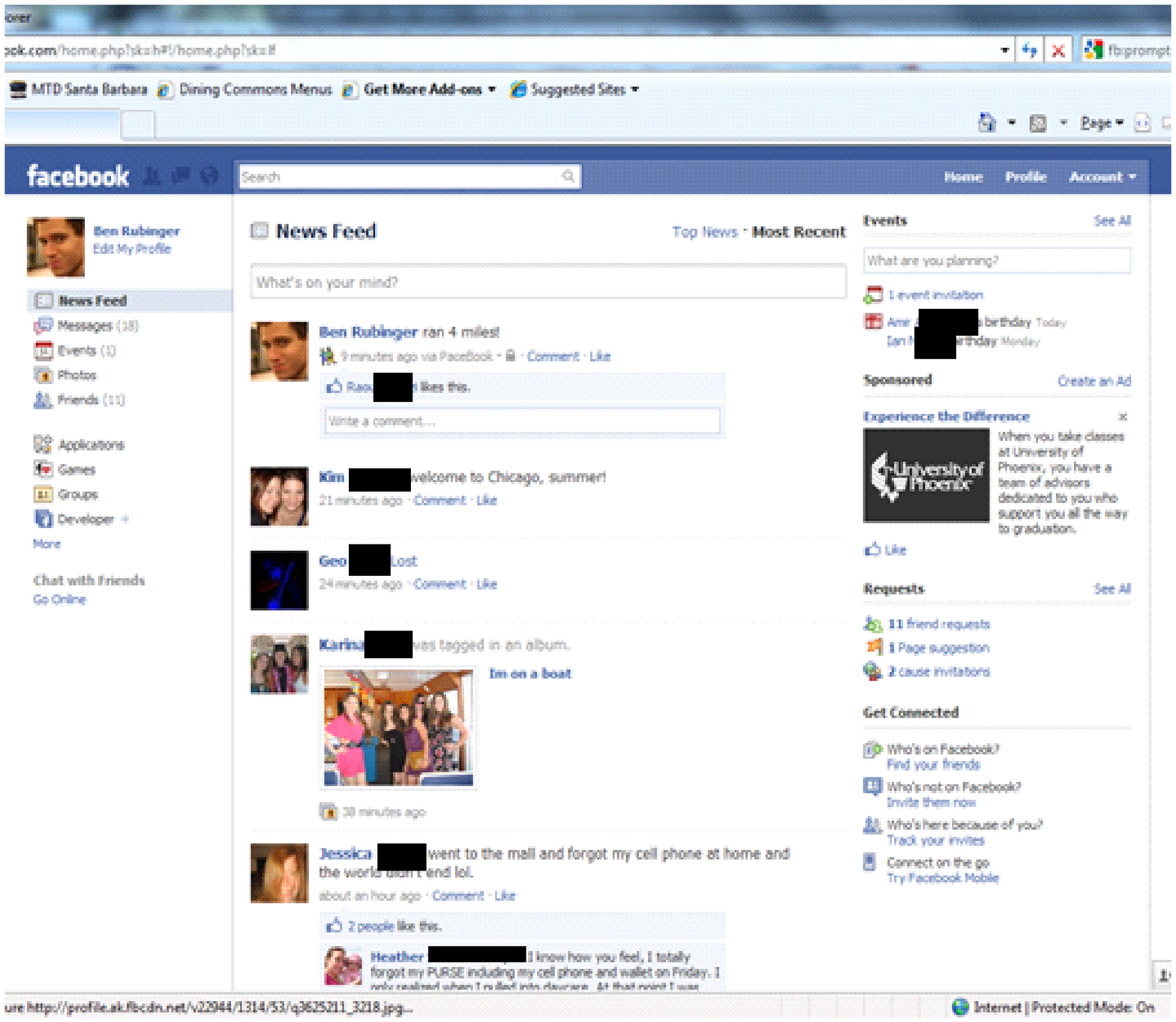}
\caption{Pacebook in newsfeed}
\label{screen2}
\end{figure}

The screenshot in Figure~\ref{screen2} shows a new entry in first author's
newsfeed which was automatically created by Pacebook announcing to all
Facebook friends of the first author of his recent run. Notice that
underneath the status update for Ben Rubinger, you see the text ``via
PaceBook.'' This is an application link allowing others to get info on
Pacebook, and ultimately add the application to their online profile so
it shows up in the left-hand navigation and they can begin using it. The
Facebook users who see the entry generated by the Pacebook can interact
with it for example  by ``liking'' it (placing a thumbs-up sign on it)
or by commenting on it.

\section{Facebook API Overview}
\label{api}

The way that a developer interacts with the Facebook system is through the
Facebook API. This API contains methods that allow developers to interact
with the application user's social data. Examples of methods in the API are:
\begin{itemize}
\item \verb!User.GetFriends()!
This returns a list of the application user's Facebook friends. Applications
might make use of this data to give the user a list of his or her friends to
choose with whom to interact in the context of a third party application. In
the context of Pacebook, this function is useful to allow the user to
``tag'' other friends that were on the run.

\item \verb!Data.GetObject(long objectid)!
This method allows objects stored in the Facebook data store (a data store
created for Facebook third party developers to persist application data)
to be retrieved.
\end{itemize}
To create a Facebook application, several different language options are available,
each having different implications. The development client library API
provided directly by Facebook company is the PHP 5 client library. As such,
applications written for this library are written in PHP and would behave
as any application running in the context of this scripting language.

There are also other options available beyond PHP for Facebook
development. One such example, the Microsoft SDK for the Facebook
Platform  allows applications written in a Microsoft .NET managed
language to integrate with Facebook. These applications can be written in
any managed .NET language, though typically the language used is either C\#
or VB.NET. For Pacebook, C\# was used.

Throughout the rest of this paper, we will refer to the Microsoft SDK for
the Facebook Platform (also known as the Facebook Developer Toolkit \cite{FacebookAPI}) as
the .NET Facebook API wrapper. The .NET Facebook API wrapper is not created
by Facebook; rather a third party consulting company called Clarity Consulting
created this wrapper in order to allow Microsoft .NET developers to write
Facebook applications. In this work we focus on writing contracts for the
.NET Facebook API wrapper. The Facebook application we use as a case study
is also written using the Microsoft .NET framework.  In our case study we
focused on the Facebook REST API in the .NET Facebook API wrapper which
contains the main functionality for accessing the social data provided
by Facebook.

\section{Microsoft Code Contracts}
\label{contract}

The Visual Studio 2010 toolset (the integrated development environment for doing
.NET development) includes new tools for performing various
analyses. These include support and tooling for using code contracts, as well
as mechanisms to perform checking against these contracts.  The contracts
are specified as pre-conditions, post-conditions, and class invariants.
The Visual Studio 2010 toolset includes support for runtime contract
checking which can be used to improve testing. The toolset also provides
support for static contract verification, and documentation generation.
Not all contracts can be checked statically, but when they can, static
identification of contract violations reduces the cost of testing, runtime
checking and debugging.

Here is an example in the context of a simple Facebook application:
\begin{verbatim}
private void TestService_Load(object sender, EventArgs e)
{
          
  ListenToEvents(true);
  try
  {
    IList<Facebook.Schema.user> Friends=null;
    Friends= FacebookService1.Api.Friends.GetUserObjects();
    FriendsHaveLongNames(Friends);
    FacebookService1.Api.GetUserPreference(300);
    ...
}
\end{verbatim}

\begin{verbatim}
private bool FriendsHaveLongNames(IList<Facebook.Schema.user> FriendList)
{
  System.Diagnostics.Contracts.Contract.Requires(FriendList != null);
  System.Diagnostics.Contracts.Contract.Requires(FriendList.Count > 5);
  return true;
}
\end{verbatim}

In the above C\# code, the method {\tt TestService\_Load} calls the method
{\tt FriendsHaveLongNames}. {\tt FriendsHaveLongNames} has two method preconditions. Both
preconditions (listed in ``requires'' calls) check a property of a ``social''
object whose value can not be determined statically and therefore neither can
the contracts. The {\tt FriendList} object passed into {\tt FriendsHaveLongNames}
is populated at runtime with social data (in this case, data pertaining
to the current application user's friend list). The
system cannot make a determination about this precondition statically. These
checks will still operate dynamically (at runtime). Should the check fail
at runtime, an exception is typically thrown (the runtime failed contract
behavior is configurable by the developer).

Separately, making the subsequent {\tt GetUserPreference(300)} call causes 
one of the contracts inside that method to fail statically as the constant integer
value can be used to check the contract statically. See the code taken from the Facebook .NET
API wrapper including the contract we have added below:
\begin{verbatim}
public string GetUserPreference(int pref_id)
{
  System.Diagnostics.Contracts.Contract.Requires(pref_id >= 0);
  System.Diagnostics.Contracts.Contract.Requires(pref_id <= 200);
  return GetUserPreference(pref_id, false, null, null);
}
\end{verbatim}


\section{Applying Code Contracts to the .NET Facebook API Wrapper}
\label{api-contract}

The .NET Facebook API wrapper source is publicly available~\cite{FacebookAPI} which allowed
us to instrument the code with contracts. 
The .NET Facebook API wrapper consists of many classes, we focused
on a small core subset that was used by the Pacebook application.
The contracts clauses were written either as a direct translation of the comments in the
API documentation or they were additional checks added by us
based on the API documentation interpretation.

\subsection{Contract Clauses: Translation from the Wrapper Documentation}

In many circumstances, the .NET Facebook API wrapper contains method
documentation explaining, in addition to other things, details about
the parameters being passed to a function. This information can be
transformed into method preconditions which specify the same information
as the documentation but do so in a way that the system is able to verify,
providing a type of enhanced and unambiguous documentation.

The following is an example of a function which is in the .NET Facebook
API wrapper source and shows the conversion of the textual documentation
into a method precondition using the ``Requires'' contract element:

\begin{verbatim}
/// <summary> Sets currently authenticated user's preference. </summary>
/// <param name="pref_id">(0-201) Numeric identifier of this preference.</param>
/// <param name="value">(max. 128 characters) Value of the preference to set. 
///     Set it to "0" or "" to remove this preference.</param>
/// <remarks>
///   Each preference is a string of maximum 128 characters and each of them 
///   has a numeric identifier ranged from 0 to 200. Therefore, every 
///   application can store up to 201 string values for each of its user.
///   To "remove" a preference, set it to 0 or empty string. Both "0" and ""
///   are considered as "not present", and getPreference() call will not return 
///   them. To tell them from each other, one can use some serialization 
///   format. For example, "n:0" for zeros and "s:" for empty strings.
/// </remarks>
\end{verbatim}

\begin{verbatim}
public void SetUserPreference(int pref_id, string value)
{
  System.Diagnostics.Contracts.Contract.Requires(pref_id >= 0);
  System.Diagnostics.Contracts.Contract.Requires(pref_id <= 200);
  System.Diagnostics.Contracts.Contract.Requires(value.Length <128);
  SetUserPreference(pref_id, value, false, null, null);
}
\end{verbatim}

\subsection{Contract Clauses: Additional Checks}

In several circumstances, we were able to insert contract clauses beyond
doing literal translation from documentation. We found that there were
instances where the documentation did not specify parameter requirements
that should have been specified. Below is an example from the Facebook
wrapper which shows a null check added as a contract clause, while there
is no mention of a non-null requirement in the documentation:

\begin{verbatim}
/// <summary>
///   Remove a previously defined object type. This will also delete ALL 
///   objects of this type. This deletion is NOT reversible. 
/// </summary>
/// <param name="obj_type">Name of the object type to delete. 
///   This will also delete all objects that were created with the type. 
/// </param>
\end{verbatim}

\begin{verbatim}
public void DropObjectType(string obj_type)
{
  System.Diagnostics.Contracts.Contract.Requires(obj_type != null)
  System.Diagnostics.Contracts.Contract.Requires(obj_type.Length > 0 );
  DropObjectType(obj_type, false, null, null);
}
\end{verbatim}

\subsection{Contract Clauses: Categories}
The contract clauses we have created for the .NET Facebook API wrapper can be divided into three categories:
\begin{enumerate}
\item {\bf Null checks:} These ensure that a null value is not passed into
a specific parameter which expects a non-null object or value.
\item {\bf Range checks:} These are used to ensure that a specified
parameter falls within a specified range. For example, there is a function,
DefineAssociation, which creates a data association between different data
types in the Facebook Data store. It has an integer parameter which specifies
the type of association (1-way, 2-way symmetric and 2 way asymmetric). It
only accepts values between 1 and 3. The contract clause created to verify
this constraint falls under the category of range checks.
\item {\bf Size checks:} These are used to verify that the length of an object
(often the length of a string or a list) meets size constraints.
\end{enumerate}

\subsection{Contracts for the .NET Facebook API wrapper}

The source code for the .NET Facebook API wrapper is responsible for
abstracting the details of the communication with the Facebook API. It
allows the developer to write code in a managed, strongly typed .NET
language like C\#, making calls which will get wrapped and sent over the
wire to Facebook's servers in XML format. It also parses the responses
sent via XML format from the Facebook server and populates strongly typed
objects for the developer to work with.

For the .NET Facebook API wrapper, there are 32 source files in the REST API, each
containing different subsections of the API wrapper. The source files we
focused on were:

\begin{itemize}

\item Data.cs (937 lines of code, 1403 lines of comments) which contains
wrapper code for interacting with the Facebook Data store (a simple, highly
scalable data store made available to Facebook third party application
developers).

\item Friends.cs (292 lines of code, 574 lines of comments) which contains
wrapper code for getting information about ``friends'' in a user's social
network.

\item Comments.cs (127 lines of code, 237 lines of comments) which contains
methods for working with comments users can post about news items that
show up in the ``news feed'' (a list of recent friend activity displayed
when the user first signs into Facebook).

\item Events.cs (243 lines of code, 517 lines of comments) which contains
methods for working with Facebook events, a type of calendar entry used
for group scheduling.  

\end{itemize}

Note that these files only contain wrapper classes that invoke the
appropriate Facebook function for each request. They do not contain
too many lines of code since they do not contain the implementation of
the functionality.

The majority of the contract clauses  we inserted to the .NET Facebook API
wrapper methods use the requires construct, i.e., the majority of the checks
we inserted were pre-condition checks.  This is the most useful check for
third party application developers since establishing the pre-condition is
the responsibility of the client and a pre-condition violation would indicate
an error in the third party application. In contrast, a post-condition
violation would indicate an error in the Facebook implementation of the API.

The Micrososft code contract system we used also includes support for
class invariants; however we have not used them in the contract we wrote
for the .NET Facebook API wrapper source code. This is because as an API,
.NET Facebook API wrapper does not afford much opportunity for making use
of class invariants since there are not many fields of relevance (because
the API maintains minimal state).

We also wrote contracts for the Pacebook application source code.
Majority of the contracts for the Pacebook application ended up in the
data access layer code, which interfaces between the user interface and the
Facebook data store.  In the contracts inserted into the Pacebook application
(rather than the .NET Facebook API wrapper source), we made use of both
class invariants  in addition to pre- and post-conditions.

The .NET Facebook API wrapper contains some public methods which in turn
call private methods which include a longer method signature. 
For example:
\begin{verbatim}
public void SetUserPreference(int pref_id, string value)
{
  System.Diagnostics.Contracts.Contract.Requires(pref_id >= 0);
  System.Diagnostics.Contracts.Contract.Requires(pref_id <= 200 );
  System.Diagnostics.Contracts.Contract.Requires(value.Length <128);
  SetUserPreference(pref_id, value, false, null, null);
}
\end{verbatim}
simply calls
\begin{verbatim}
private void SetUserPreference(int pref_id, string value, bool isAsync, 
                       SetUserPreferenceCallback callback, Object state)
{
  var parameterList = new Dictionary<string, string> { 
    { "method", "facebook.data.setUserPreference" } };
  Utilities.AddRequiredParameter(parameterList, "pref_id", pref_id);
  Utilities.AddRequiredParameter(parameterList, "value", value);
  if (isAsync)
  {
    SendRequestAsync<data_setUserPreference_response, bool>(parameterList, 
                        new FacebookCallCompleted<bool>(callback), state);
    return;
  }
  SendRequest(parameterList);
}
\end{verbatim}
Since the developer interacts with the public method, that is where
we placed the contracts as this avoids redundant checks.

The .NET Facebook API wrapper also includes asynchronous versions of
some methods where the method call returns immediately without finishing
the task.  The task is executed on a separate thread and the caller is
notified of completion by an event. This style of asynchronous interaction
avoids blocking of the calling thread. We have not written contracts for
this type of asynchronous methods in the API since they were not used in
the Pacebook application.

\begin{table}[htdp]
\centering
\begin{tabular}{|l||c|c|c|c||c|}\hline
   & Data.cs & Friend.cs & Comments.cs & Events.cs &  Total \\  \hline\hline
\# Methods & 39 & 10 & 6 &  11 & 66 \\
\hline
\# Contracted & 33 & 8 & 6 &  9 &  56 \\ 
\hline 
Percentage & 92\% & 80\% & 100\% & 82\%  & 85\% \\ 
\hline
\end{tabular}
\caption{Number of methods (and contracted methods) in Facebook API classes we investigated}
\label{methods}
\end{table}%

\begin{table}[htdp]
\centering
\begin{tabular}{|l||c|c|c|c||c|c|c|}\hline
   & Data.cs & Friend.cs & Comments.cs & Events.cs & Total & Contracted & Percentage\\  \hline\hline
\# Parameters & 92 & 17 & 17 & 21 & 147 & 101 & 69\% \\
\hline
\# Nullable & 60 & 7 & 11 &  16 & 94 & 76 & 81\% \\
\hline
\# Range-able & 79 & 11 & 17 &  7 & 114 & 32 & 28\% \\
\hline
\# Size-able & 55 & 4 & 11 &  5 & 75 & 26 & 35\% \\
\hline
\end{tabular}
\caption{Number of method parameters (and contracted method parameters) in Facebook API classes we investigated}
\label{parameters}
\end{table}%

Tables~\ref{methods} and \ref{parameters} provide some statistics about the contracts we
wrote for the Facebook API.
Table~\ref{methods} gives the number of methods in the four files of the Facebook API that we have investigated.
For each of them, we also specify the number of methods that we wrote contracts for. 
Overall we wrote contracts for 85\% of the methods. For the rest of the methods there were no
conditions described in the API documentation and we were not able to find interesting contract clauses to
write.

In Table~\ref{parameters} we provide data about the number of parameters for the methods we investigated
in the Facebook API. We categorized the parameters as parameters that can be constrained with
null (nullable), range (range-able) or size (size-able) constraints respectively. Note that these
categories are not mutually exclusive. For example a list can have both a size constraint on it identifying
a maximum size and a null constraint which may state that it should not be null.
In Table~\ref{parameters} we also identify the percentages of each type of parameters that were
constrained with a contract clause. Not surprisingly, the most common contract clauses were on
null values. The percentages for range-able and size-able parameters are relatively low  but this
should be expected since, for example, although any integer parameter can be range-able most of them
may not have a specified range in the API specification. 
As seen in the first row of Table~\ref{parameters}, overall, 69\% of the parameters had some
type of contract clause written about them, restricting their possible values.

\section{Discussion}
\label{discussion}

The process for searching for bugs in the code that interacts with the
Facebook API is often cumbersome. Code is created or modified by the
developer, and then to test this code, it is deployed to the web server,
and then the web site must be accessed via a web browser to execute
it within the context of the server through the Facebook website. This
is necessary because the code will be calling the Facebook API and this
additional runtime dependency adds complexity to the debugging process. The
developer then can see if there is an unexpected runtime behavior such as
an exception being thrown to identify the bugs.

Let us contrast this experience with the new experience we achieve with
the modified source code which includes the code contracts.  For example,
if we look at the method {\tt GetUserPreference}, we can see that the API
has a parameter for id of the stored preference. This parameter can only
accept the values from 0 to 200.  If a developer were to be unaware of
this restriction or make a mistake and use an invalid value, to discover
this error, he or she would go through the following process:
\begin{itemize}
\item Upload the new code to the web server
\item Attempt to view the corresponding web site
\item View the relevant exception which was thrown.
\end{itemize}
This is a time consuming and cumbersome multistep process. Conversely, with the
static contract checking features in the Microsoft code contracts system
paired with the enhanced contracts-added version of the .NET Facebook
API wrapper source code, the developer is prompted that she/he made a
coding error after typing in the wrong value for the input parameter
and the violated contract clause is highlighted to inform the developer.
This process is of course dependent on the system's ability to statically
determine the value that will be used for the specific parameter. Even
if the value cannot be determined statically, during runtime the contract
system will inform the developer about any contract violation and identify
the contract clauses that are violated, which significantly improves the
efficiency of the debugging process.

In order to track the usefulness of the contracts we inserted, throughout
the development of the Pacebook application we tracked the number of defects
that were caught by the contract system.  There were  eight defects that were
caught with the aid of the inserted code contracts. Of those defects, six
were caught by contracts in the .NET Facebook API wrapper source, while
the other two were caught by contracts in the Pacebook application source.
Of the 6 defects caught by the contracts in the .NET Facebook API wrapper
source 4 were null check violations, 1 was a range check violation and 1
was a size check violation.  The defects caught by the contracts in the
Pacebook application were both null check violations.

It is important to give some context regarding the above numbers. The
code contracts were inserted into the source in an early stage of the
development of the Pacebook application. Accordingly, the developer for
the Pacebook (the first author) had reviewed the documentation,
and had essentially retyped the documentation in contract form which made
him familiar with the documentation and the meaning of the newly inserted
contracts. Having this knowledge made the developer less likely to make
some of the errors that otherwise would have been more likely. For others
who develop their Facebook applications with the newly instrumented .NET
Facebook API wrapper, it is reasonable to expect even greater value from
the contracts since they will not have the same recent familiarity with
the inserted contracts.

Let us now focus on a specific bug that was caught with the aid of the
contracts. A call in the Pacebook application was unintentionally passing in
an invalid value for a parameter of the method {\tt CreateAssociation}. With the
contracts plus static checking, this was caught before deployment without
a need to go through the cumbersome bug finding process described above.

Another interesting bug that was discovered by contract checking was a bug in
the contracts themselves. The static checker showed that a contract clause
had failed for code that was not being called by the Pacebook application,
but rather was being called by the API itself. It was then realized that a
null check was inserted where a null parameter was acceptable. Note that this
types of bugs are also very valuable to find since they show the inconsistency
between the API description and its implementation. In contrast, a similar
erroneous statement written as a comment as part of the API documentation
cannot be discovered by contract checking and can lead to misunderstanding of
the API by the third party developers. Precisely specifying the assumptions
about the API using contracts enables us to check the consistency between
the API implementation and its specification using contract checking.



\section{Conclusions}
\label{conclusion}

Our experience shows that code contracts are a valuable tool for the
development of third party Facebook applications. The contracts we have
written for the Facebook API clarify the assumptions about how to interact
with the Facebook API and, hence, are likely to reduce the number of errors
in third party applications. We have observed that static contract checking
is a useful tool for identifying the errors at compile time instead of the
cumbersome manual testing process that requires the developer to launch the
application. We believe that the contract we started for the Facebook API
can be extended to the full API and can serve as a precise documentation
of the API for the third party developers.

\bibliographystyle{eptcs} 
\bibliography{bultan}

\end{document}